# The 1908 Tunguska Event And The 2013 Chelyabinsk Meteoritic Event: Comparison Of Reported Seismic Phenomena

Andrei Ol'khovatov

https://orcid.org/0000-0002-6043-9205
(Retired physicist)
Russia, Moscow
email: olkhov@mail.ru

**Dedicated to the blessed memory of my grandmother ( Tuzlukova Anna Ivanovna ) and my mother ( Ol'khovatova Olga Leonidovna )**

**Abstract:** Till now there is no solid scientific proof for any of the numerous hypotheses proposed for the Tunguska 1908 event. Regarding the most popular "spacebody infall" interpretation it is reasonable to compare the Tunguska event with the 2013 Chelyabinsk meteoritic event which energy deposition was about a half megaton TNT. In this paper seismic manifestations are compared. The comparison demonstrates that while a "seismic image" of the 2013 Chelyabinsk meteoritic event is an agreement with the "spacebody infall" theory, the 1908 Tunguska event one strongly differs. There are evidences of a swarm of earthquakes on June 30, 1908 in the region of the Tunguska event manifestation.

## 1. Introduction

More than a century has past after the famous 1908 Tunguska event in Siberia (abbr. Tunguska). But till now there is no solid scientific proof for any of the numerous hypotheses proposed. Regarding the most popular "spacebody infall" interpretation it is reasonable to compare Tunguska with the 2013 Chelyabinsk meteoritic event (abbr. 2013CME) which energy deposition was about a half megaton TNT, while for hypothetical Tunguska spacebody fall is now considered to be just 3-5 Mt TNT [Boslough, 2013], i.e. ~6 - 10 times more energetic.

Seismic stations data of the 1908 Tunguska event was analyzed in several papers by A. Ben-Menahem [Ben-Menahem, 1975], and I. P. Pasechnik [Pasechnik, 1976, 1986]. Both authors analyzed data of the seismic stations and came to similar conclusion that amplitude of seismic oscillations detected by the stations were equal to be produced by an earthquake with magnitude M ~ 4.5 - 5.

However a little analysis was done on seismic phenomena reported by witnesses. The Irkutsk seismic station had a network of correspondents in the region who regularly informed the station about seismic phenomena felt in their places. And there were reports on the events of June 30, 1908. This article will compare the testimony of the witnesses reports with the associated station data both during the 1908 Tunguska event and 2013CME.

Translation from Russian of original Tunguska witness's accounts and some other texts is done by the author of this paper (i.e. A.O.), unless otherwise is stated. The translation was done as close to originals as possible. Translations of large volumes (as well as citation in English) are marked by cursive font. In some places of translations an alternative translation and/or comments in { ...} are given by the author of this paper for better understanding. Please pay attention that the Tunguska event occurred on June 30, 1908 (on the Gregorian calendar) or on June 17, 1908 on the Julian calendar which was widely used in Russia in 1908.

## 2. The 2013 Chelyabinsk event

2013CME was seen (and/or videoed) up to distances ~700km and maybe even a bit farther. While infrasound was detected by many monitoring stations all over the world, audible sounds were reported only at much closer distances. The farthest report about sounds is from the settlement of Sosnovka at the distance 171 km [Popova et al., 2013]. This is in agreement with observations of other bolide's events where audible sounds usually are reported at distances up to ~ 200-300 km.

Remarkably that electrophonic phenomena accompanied the 2013 Chelyabinsk meteoritic event [Popova et al., 2013] are similar to those which accompany meteoroid's entries with energy deposition several orders of magnitude less. This is in favor to the author idea that such energy deposition is sooner a trigger of some processes which we don't understand still (see in [Ol'khovatov, 2021a]).

Estimations of the seismic event's magnitude caused by 2013CME vary a little bit from one research to another, but most of them give about Ms ~4.0 or a little bit less [Heimann, et al., 2013]. A remarkable feature of the generated seismic waves is [Heimann, et al., 2013]:

*"Unlike tectonic earthquakes or underground explosions, the body
waves are almost absent, and sharp onsets cannot be identified
even at the closest station, ~220 km away. Almost no seismic
energy is detected above 0.1 Hz, or on transversal components,
in contrast to tectonic earthquakes with similar seismic moment."*

In other words, if to exclude an area close to the epicenter, then the propagating seismic waves consist almost of low-frequency surface waves. For witnesses this means slow "swaying" and not sharp "jerking/jolts".

The author of this paper discovered in [Popova et al., 2013] that the largest distance with witness's reports about seismic was 96 km. This is maybe a little bit small for M=4 earthquake, but anyway does not strongly deviates from the magnitude's estimations (see also below). The slow "swaying" can be more easily unnoticed than the sharp "jerks/jolts".

Behavior of seismic waves produced by a large meteoroid in many aspects is similar to the waves produced by powerful aerial explosions. Attention was drawn to this similarity back in the mid-1970s [Ben-Menahem, 1975; Pasechnik, 1976 ], a long time before 2013CME. So the seismic phenomena associated with 2013CME and reported by witnesses are in accordance with theory, and do not deviate from the row of other numerous meteoroidal bolide's events despite its large energy.

## 3. The 1908 Tunguska event

In Tunguska some "luminosity" was reported up to distance 710 km [Krinov, 1949]. But reports about hearing sounds and earthquake came from much larger distances. Ye. L. Krinov marks in [Krinov, 1949] that the farthest place where sounds phenomena (compared with several cannon's shots) were noticed is Achaevskii ulus positioned about 1200 km to SW from the Tunguska epicenter (here the Tunguska epicenter (the epicenter) means the "center" of the Kulikovskii forestfall calculated by W.H. Fast with his colleagues, i.e. about 60.9° N, 101.9° E [Pasechnik,1986] ).

According to Krinov the farthest seismic phenomena were reported from Zhymygytskii stan positioned 1025 km to the south from the place of the epicenter. Krinov also presents a dozen other reports about seismic phenomena at distances ~700 - 1000 km.

Another prominent researcher of Tunguska - I.S. Astapovich wrote in [Astapovich, 1951] that the earthquake was felt over a square of a million square km., and  he placed isoseismals of intensity I=4 (Rossi-Forel scale) about 900-1000 km from the Tunguska epicenter.  So the Krinov's and Astapovich's conclusions are similar.

Astapovich also discovered a tendency to stronger seismic effects to the south from the epicenter direction, which he explained as influenced by a "meteorite's ballistic wave" (when Astapovich wrote his paper  it was thought that the "meteorite" flew from the south to the north, while in the mid-1960s it was changed to "almost from the east to the west").

Here is a map of the earthquake by Astapovich [Astapovich, 1951] on Fig.1, where the dotted lines show isoseismals, the AM line is the assumed trajectory of the Tunguska spacebody, and Lake Baikal is in the lower right corner.

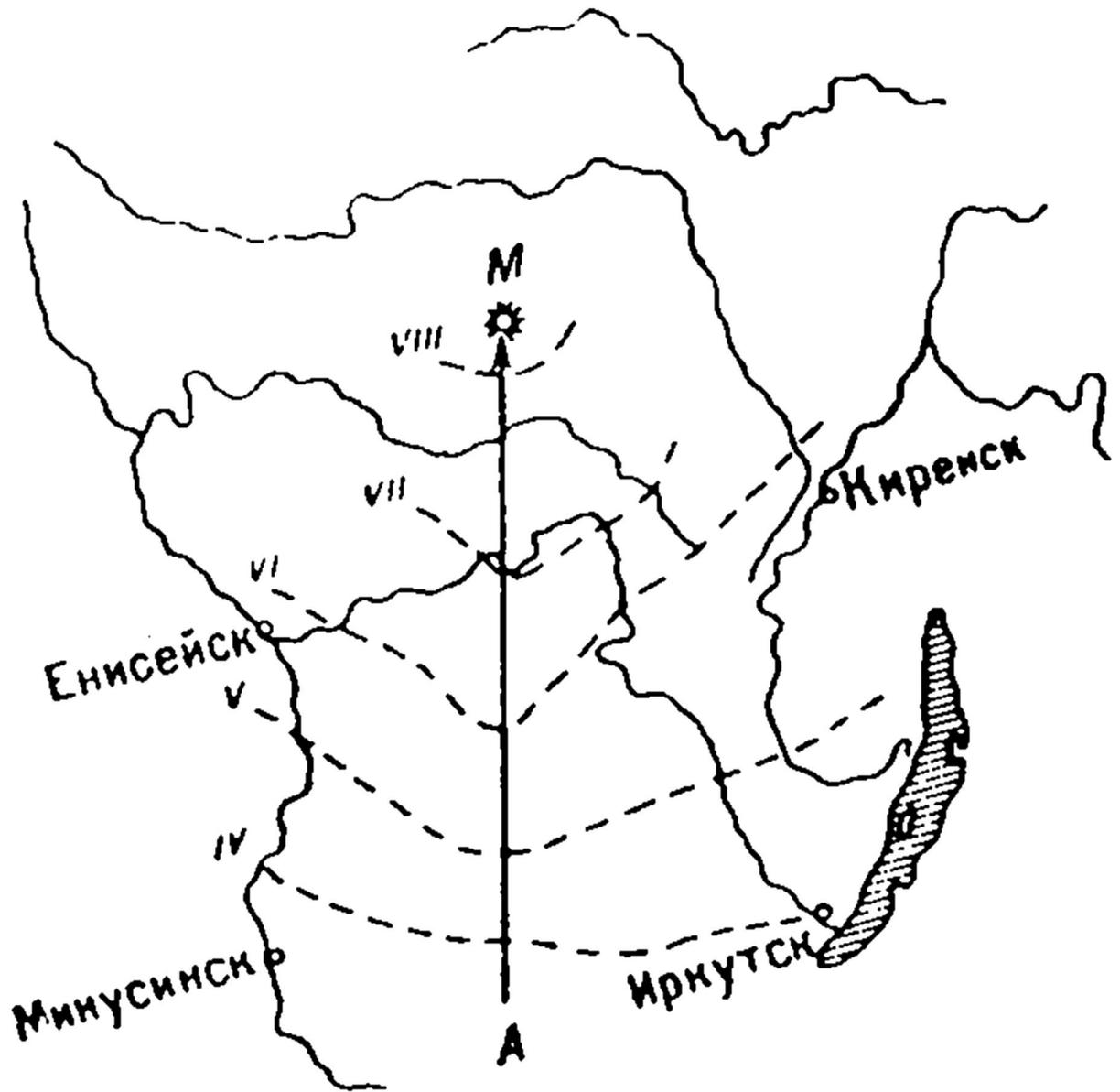

**Fig.1**

It is noteworthy that the isoseismals of intensity I=4 (Rossi-Forel scale) on Fig.1 pass almost through Irkutsk, where the earthquake appeared as just a low-amplitude record on a seismogram. The possible reason for this will be discussed below.

One of the major sources for the works by Krinov and Astapovich were witness's accounts collected by A.V. Voznesenskii who in 1908 was a director of the Magnetic-meteorological observatory in Irkutsk (for shortness it is often called just as "seismic station" in this paper). A.V. Voznesenskii had a large network of correspondents in the region who regularly reported him about earthquakes. Voznesenskii had got ~60 reports about the "Tunguska earthquake" in total. It is very important that the reports were delivered on "hot traces" after the event and by rather experienced observers.

Here are just several such reports most of which are taken from [Vasil'ev et al., 1981] unless otherwise stated. Also a distance from the Tunguska epicenter and an azimuth from there is given mostly from [Vasil'ev et al., 1981] and [Razin, 2008] . Translation from Russian is by the author of this paper (i.e. A.O.).

Voznesenskii also spread a questionnaire (let's call it as: questionnaire A) regarding: a place of observation (1); a date and time of the observation (2); details of the earthquake and whether it was accompanied with subterranean rumble (3); about possible damage and other effects of the earthquake (4); about an intensity of earthquake (5); surname and address of the observer (6). Dates in the accounts are in the Julian calendar unless otherwise is stated. Here are translations of some of the collected accounts:

"

*1. Plaksin I., a letter of June 22, the settlement of Tangui, the Bratsk district, the Nizhneudinsk country {613 km, 185° -A.O.}*
*2. 17 June, about 8 a.m. on Tuesday.*
*3. Only one jerk {a "push" –A.O.} was noticed. Tottering {"staggering"' –A.O.}*
*4. It did not produce.*
*5. Practically unnoticed.*
"

The next account is with some luminous phenomenon (translated):

"

*L.D. Klykov - a chief of the post-office in the settlement of Znamenskoe,  the Irkutsk province {707 km, 164° - A.O.}, July 11, 1908.*
*2. June 17, 1908 on Tuesday about 8 a.m. (watches were verified with the Zhiganovskii telegraph )*
*3. Subterranean rumble was not heard... There were two jerks { "pushes"-A.O. } with time difference one from another 3-4 seconds. The whole earthquake lasted 7-8 seconds.*
*4. It was not. Simultaneously with it a fall of an aerolite was seen: approximately to the south-west from the settlement of Znamenskoe; a fiery strip was seen, a sky was completely clear, a strong thunder was heard with an explosion at the end.*
"

Interestingly that the fiery strip was seen in the direction where the hypothetical Tunguska spacebody could not fly.

Voznesenskii also spread a new more detailed questionnaire (let's call it: questionnaire B) about the June 17 earthquake including the questions:
1. A place of observation;
2. A time of the beginning and the end of the earthquake;
3. Direction;
4. Details of the quaking;
5. Was a rumble heard?
6. Personal impressions.
7. Was furniture, utensils moved?
8. Here Voznesenskii asked to evaluate the earthquake by Rossi-Forel scale and explained the scale.

Here is an interesting and important reply. A priest D.A. Kazanskii from the settlement Zhimygytskii stan (1010 km, 176° from the epicenter) reports (translated):

*"1. Stan Zhimygytskii, in the yard.*
*2. 8 a.m., the end at 8 h. 20 min. a.m.*
*3. The north - west.*
*4. Slight shivering.*
*5. There were like rolls of thunder, and then cannon's shots far away.*
*6. I suggested mountain rocks falling.*
*7. It was not.*
*8. 3 (and underlined "majority") { i.e. "it was felt by majority of people – A.O.}.*
"

A chief of a meteorological station in the settlement of Maritui wrote to Voznesenskii on July 26, 1908 about the following accounts collected by him, including this one (translated):

"

*The railroad watchman Alekseev came from duty about 8 a.m. (he does not have watches) into the room on the verst 78 of the Trans-Baikal Railway, near the "Tolstyi mys" on Lake Baikal, and was going to drink tea. At that time, he noticed that the lantern was swinging and the icon on the shelf near the wall facing south fell on the floor. The oil from the lantern spilled out. Swing of the lantern was about 1/4 of arshin {~18 cm - A.O.}. The lantern swayed in the direction of the NNW to the SSE. In the same direction, the icon fell from the wall. Soil oscillations and noise were not noticed. The marked phenomena were observed by Alekseev himself, his wife and two children.*

"

The place "Tolstyi mys" is 1023 km from the Tunguska epicenter at azimuth 169° .

One more important account collected by the chief of the meteorological station in the settlement of Maritui (translated):

"

*Tunnel watchman Shinelev about 8 o'clock in the morning, being on duty on the verst 110 "Troinaya guba" heard a noise as if from under the ground. Soon he noticed quite significant screes of small stones from a slope on the way. He did not feel oscillations of the soil.*
"

The place "Troinaya guba" is 1019 km from the Tunguska epicenter at azimuth 170 degrees.

There are several interesting reports from the region of golden mines. Here is a scan (Fig.2) from the "Sibirskaya Zhizn" newspaper of August 14, 1908 (Julian calendar):

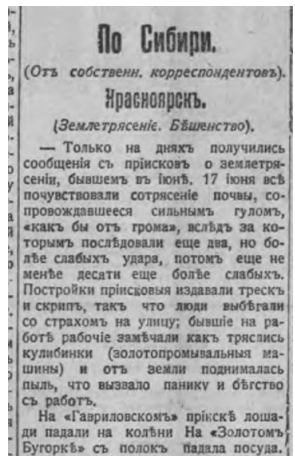

**Fig.2**

Here is the translation of the newspaper note:

*"Only the other day there were reports from the mines  about an earthquake that occurred in June.  On June 17, everyone felt a shaking of the ground, accompanied by a strong rumble, "as if from thunder," followed by two more, but weaker blows, then at least ten more weaker ones. The mine buildings made a cracking and creaking sound, so that people ran out into the street with fear; workers*

who were at work noticed how the kulibinki (gold washing machines) were shaking and dust rose from the ground, which caused panic and runaway from work.

　　*At the Gavrilovskii mine, horses fell to their knees. Dishes were falling from the shelves on the "Zolotoi bugorok" {mine – A.O.}."*

The Gavrilovskii mine was about 493 km from the epicenter at azimuth 266° , the Zolotoi bugorok mine was about 493 km from the epicenter at azimuth 267° .

　　A couple of more the 1908-reports was presented (translated in English) by Francis John Welsh Whipple [Whipple, 1930]. He took the reports from the Bulletin of the Permanent Central Seismic Commission of Russia (for 1908). Here is his translation:

　　*"June 30. Town of Kansk.—The first shock caused the doors, window and votive lamp to shake. Subterranean rumblings were heard. About 5 to 7 minutes later there was a second shock accompanying the rumbling. A minute later there was a further shock less severe than the preceding two. It is stated that the earthquake was accompanied by the fall of a meteorite near the village of Dalaia. Peasants relate that 70 km. north of Kansk in the Ustianovski district there was also an earthquake accompanied by subterranean rumblings.*
　　*June 30. Kuriski-Popovich Village; District of Kansk.—At 0 h. 37 m. a severe earthquake was observed in the vicinity of the village. After this there were two loud bursts, like the firing of a large calibre gun near Lovat Village. It was afterwards found that a large meteorite had fallen."*

Kansk is 630 km at azimuth 217° from the epicenter, the village Dalai is about 15 km to NE of Kansk, the village Kurysh-Popovichi is about 15 km to the north of Kansk. The village Lovat' is about 10 km to NE of Kansk.

　　There were other reports from the region of gold mines right after the event. For example,  "Bulleten' Postoyannoi tsentralnoi seismicheskoi komissii" (the Bulletin of the Permanent Central Seismic Commission) of Russia for April-June 1908 contains a message from the golden mine Stepanovskii of the South Yenisei Mountain District  that on June 29 at 23 h 43 min GMT  there was heard for several seconds a strong underground rumble, like three cannon shots.  The accompanying vibrations of the soil shook wooden buildings throughout the area occupied by the basins of three rivers: Uderei, Borovaya and Murozhnaya. The Stepanovskii mine is situated at about 488 km from the epicenter at azimuth about 243° . It is remarkable how this report is discussed in [Whipple, 1930]:

*"In the Bulletin of the Permanent Central Seismic commission
a report by a certain Peter Sukhodaeff appears under the date
June 29, whilst the summaries of newspaper reports quoted above
are given under June 30. The only times stated explicitly are
June 29, 23 h. 43 m. G.M.T. for " a subterranean rumbling,
like three rounds from a cannon, heard in a period of a few seconds "
at the Stepanovski Mine, South Yenesei and June 30, 0 h. 37 m.
G.M.T. for a severe earthquake observed near Kansk.*

*Evidently a mistake of an hour has been made in the former
estimate of the time or in the conversion to G.M.T. The disturbance which was
propagated to the seismological stations
originated at 0 h. 15 m. G.M.T. and the supposed earthquake was
felt by the inhabitants of the region near Kansk some 30 minutes
later. As the distance from the falling place of the meteor to
Kansk is about 600 km. the time is appropriate for air waves and
we are justified in assuming that the reporters were misled in supposing that they had
felt an earthquake. Any disturbance which
was propagated through the ground had passed by more than
twenty minutes earlier.*

*In the report from the town of Kansk (see above p. 288) it is
stated that the first shock caused the doors, windows and a votive
lamp to shake and that there was an interval of 5 to 7 minutes
before the second shock. It is not impossible that the first shock
was actually seismic. A lamp might well be set swinging by the
earth movement. The interval of 5 to 7 minutes was in that
case badly underestimated. There is, however, no suggestion of
any long break in the phenomena in any of the other reports from
Kansk quoted above. On the whole the evidence is against the
hypothesis that any wave through the ground was felt at Kansk."*

Here are just two remarks:

1) one hour seems to be too large for a mistake. Local time in 1908 was calculated from geographical longitude (i.e. 1 degree is equal to 4 minutes of local time). Moreover the Stepanovskii mine was a kind of a central office of several golden mines. In general golden mines were rather well-equipped (some of them even had steam engines and local phone's lines in the mid of the wild taiga).

2) The reported seismic manifestations in Kansk are well-known during earthquakes and caused by "waves through the ground", and are similar to those already presented in this paper. Anyway here are some more examples. The observer (T. Grechin) of the Shamanskie water-measuring posts wrote to Voznesenskii on June 18 i.e. on the next day after the event (translated):

*"On June 17, at 8 o'clock in the morning, there were some strong blows several times in the north side like thunder, from which the windows trembled in the frames, the trees bent down and the leaves of them shook; at the same time it was clear and quiet, and even the water did not lose its gloss, no damage was noticeable. As local peasants who were at field work at that time told me, they saw some kind of a fireball flying in the north side, from which such strong blows like explosions seemed to occur."*

"The trees bent down" without any noticeable air-action is a very remarkable fact. The Shamanskie water-measuring posts are situated about 586 km at azimuth about 180° from the epicenter.

Let's look for a comparison at description of earthquake occurred in 1856 near the town of Kirensk [Petrov, 1857], which is positioned at 492 km, and azimuth about 132° from the epicenter. Here is a witness description (translated):

*"Suddenly, a deafening explosion broke out in the western side; it was followed by more blows, which finally merged into the roar of a huge collapse, the rumble from which, having been heard in the mountains, lasted for several seconds and resembled a hurricane that had begun. At the very moment of the explosions, first we felt a strong jolt, and then an extraordinary shaking of the earth, which lasted for almost half a minute. At this time, the water on the lake was agitated and trees and grass were swaying around us. It must be confessed that the suddenness of this extraordinary phenomenon brought upon us an incomprehensible, one might say, painful fear. The rotation of the earth was so strong that we could barely stand on our feet; but the cattle walking near us roared furiously, some fell, another spread his legs to stay on the ground. Then, little by little, the echoes of this rumble subsided, and the blows were no longer repeated.*
*The same explosions were heard in Kirensk itself, but there was no damage. The buildings were shaking, but no harm was done. In the surrounding villages, at a distance of 5, 7 and 12 versts, the same phenomenon was happening, and, by the way, in the village of Chernoslobodskoye, the shaking was so strong that windows fell out in many houses. The peasants and shepherd boys who were in the field certify that before the underground rumble and earthquake they saw on the other side of the Lena*

*River opposite their village, behind a steep red sandstone cliff, in the taiga valley a strong light, as if fiery, and then there were explosions and shaking of the earth with a strong noise. One of the shepherds adds to this testimony by the fact that when the blows broke out, at that moment the clouds hanging over the mountain suddenly drifted away.*

*Later I asked the postman who accompanied the mail from Irkutsk to Kirensk about this phenomenon: he reported that on the road, 125 versts from the town, he heard two blows and an earthquake; the same was confirmed by the peasants who lived below Kirensk on the Lena, 50 versts away."*

Interestingly that in 1930s some researchers of the 1908 Tunguska event thought that probably a large meteorite fell near Kirensk in 1856, as the description (especially the "strong light") resembled accounts of the 1908 Tunguska event! However nowadays the 1856 event is surely assigned to a tectonic earthquake [Radziminovich, Shchetnikov, 2008].

In the annals of the Irkutsk seismic station (the Observatory) the "1908 Tunguska event" earthquake received the number 1536. In 1925 Voznesenskii published an article [Voznesenskii, 1925] in which, in particular, he wrote (translated):

*"It should be noted that a little earlier (namely at 18 h. 44 m. Greenwich time on June 28) not only by all devices of the Irkutsk Observatory, but also the instruments of the stations in Chita and Kabansk was noted a stronger earthquake No. 1535, and therefore the Observatory, according to the procedure adopted for studying local concussions, immediately sent out survey forms to all its correspondents, with a request to report everything they know about the earthquakes on June 29 and 30.*

*With the answers received, it was established quite reliably that earthquake No. 1535 was noticed by many people to the east of Irkutsk; then earthquake No. 1536 was noted by very few people, mainly in the southern strip of Siberia on June 30, as a ground shaking, but a significant majority of respondents indicated that they did not notice any ground shaking on June 30, but sounds most similar to cannon shots from large guns or a very strong thunder were clearly audible both to them personally and to other persons. All this happened around 8 a.m. local time with a completely cloudless sky. Some correspondents in the northern part of the region noted clear light phenomena and definitely talked about a giant meteor that shone brightly, the fall of which was accompanied by a number of loud sound phenomena. Correspondents of various Siberian newspapers also described it quite consistently."*

Let us note that in those years, a meteor often meant (especially to ordinary people) not only glow due to a spacebody entry, but many other luminous natural phenomena.

Voznesenskii tried to summarize the information received [Voznesenskii,1925] (translated):

*"The same kind of information was received by us from various correspondents, numbering over 60.*
*<...>*
*As can be seen from this list, by almost all observers strong acoustic phenomena are observed, at a huge distance between Yenisei, Lena and Baikal. Extreme points in the north — Upper-Inbatskoye and Mukhtuya, in the south—a number of points on railroad, starting from Kansk to Baikal. <...>*
*There is much less information about light phenomena. They are noticed only in 17%. This is explained firstly by the fact that the weather was clear in the south of the affected region — in the north it was cloudy, and secondly, for a good half of all observations, the meteor obviously flew close to the Sun, so it was accessible to the observer only in the eastern part of the region. All 17% observers of the light phenomena were located in the eastern part of the region.*
*30% of observers note more or less strong concussions, if not always concussions in the full sense of the word, of the soil, then, perhaps, trembling from a sound wave."*

The closest to the epicenter and "official" report with details of the event and dated 1908 is from Kezhma, which was (now abandoned) at distance 217 km from the epicenter at azimuth 192° . In Kezhma A. K. Kokorin (an observer of the Kezhma meteorological station) wrote in the meteorological observations log for June 30, 1908 (translated):

*"At 7 a.m., there appeared in the north two fiery circles of colossal dimensions; 4 m. after their appearance, the circles vanished; soon after the disappearance of the fiery circles, a powerful noise was heard, resembling the sound of wind, which went from north to south; the noise lasted about 5 m. Then followed sounds and crashes resembling discharges of giant cannons, which made the windows rattle. These shots continued for 2 m. and then a crackling was heard, resembling rifle shots. The latter continued for 2 m. All this happened when there was a clear sky."*

Interestingly that there is no mentioning about giant bolide/fireball! The observer - Anfinogen Kesarevich Kokorin (1872 – 1938) was interested in the event, collected info, etc. In the late 1920s he even helped to L.A. Kulik (Kulik wrote in 1930 in a letter that Kokorin was considered as a valuable employee, but was "lishenets" – i.e. deprived of some civil rights).

The Kokorin's report does not mention any large ground oscillations, despite being much closer to the epicenter than places of other 1908-reports.

It is interesting to note that in 1960s V.G. Fesenkov (academician of Academy of Science of USSR, and the chairman of the Committee on Meteorites) tried to calculate the "Tunguska spacebody" trajectory basing on the early eyewitness accounts about the light phenomena, and he had to wrote such a phrase [Fesenkov, 1966] (translated):

*"These are rather uncertain conclusions that may be inferred from a review of the most trustworthy eyewitness accounts of 1908."*

The Fesenkov's failure to get a reliable trajectory hints that possibly the Tunguska event was associated with more than one luminous body/phenomenon, but discussion of this is beyond the scope of this paper.

In this paper we concentrate on reports/accounts of 1908, as those taken many years later are less reliable. It is noteworthy that the most recent message from 1908 about the event (at least from those known to the author) was received in the mid-1970s. A local historian E. Vladimirov found in the archives of the city of Krasnoyarsk the report of the Yeniseisk's police officer to the governor dated June 20, 1908. The local historian wrote about his find in the newspaper "Krasnoyarskii rabochii" (N 186, August 10, 1975). Here is on Fig.3 a scan of a fragment with the report from this article:

июня 1908 года.

«В селе Бельском, Бельской волости, — говорится в рапорте, — в 7 часов утра 17 июня жителями и становым приставом Кошелевым было замечено сотрясение зданий с дребезжанием стекол, повторившееся дважды через короткий промежуток времени при совершенно ясной погоде и отсутствии грозовых туч и даже облаков. Повреждений

или несчастий от этих подземных толчков не произошло...»

**Fig.3**

Here is translation of the report presented in the 1975-article:

*"In the village of Bel'skoe, the Bel'skii district, at 7 o'clock in the morning on June 17, residents and police officer Koshelev noticed a shaking of buildings with rattling glass, repeated twice in a short period of time with absolutely clear weather*

*and the absence of thunderclouds and even clouds. There was no damage or misfortune from these underground shocks..."*

The village Bel'skoe is situated 650 km from the epicenter at azimuth 243° .

The seismic disturbances were also recorded by 4 seismic stations: Irkutsk, Tashkent, Tiflis, Jena [Ben-Menahem, 1975; Pasechnik, 1976]. The Irkutsk one was the closest one to the Tunguska epicenter (about 970 km to the south from the epicenter). Extremal peak-to-trough of the seismograph recording in Irkutsk was 4 mm.

Ben-Menahem [Ben-Menahem, 1975], and Pasechnik [Pasechnik, 1976] based their analysis on the assumption, that the seismic recordings were caused by a spacebody explosive destruction over the epicenter, i.e. dimensions of the source of the seismic waves were small compared with the distances (short-term point-like source). In frames of this assumption they came to conclusion that the registered Tunguska event's earthquake had a magnitude Ms~5.0 [Ben-Menahem, 1975] or ~4.5…5.0 [Pasechnik, 1976], and occurred at about 00.14 GMT, June 30 (the Gregorian calendar).

## 4. Discussion

Remarkably that spreading of the Tunguska seismic phenomena strongly differs from the 2013CME. In 2013CME witness's accounts agree with the seismic station data, but in Tunguska they disagree. Indeed for the Baikal region there is an equation for estimation of seismic intensity (I) at various distances from the epicenter R (km) of an earthquake with magnitude M (based on surface waves) [Radziminovich, Shchetnikov, 2011]. For the near surface source the equation simplifies into:

$$I = 1.5 \cdot M \ - \ 4 \cdot \lg(R) \ + \ 4$$

It can be seen from the equation that an earthquake with magnitude 4.5 – 5 can be felt within ~300 - 400 km from its source. A good illustration of this result is a recording at the Magnetic-meteorological observatory in Irkutsk where the "Tunguska earthquake" was detected only as low-amplitude ground oscillations by a seismometer [Ben-Menahem, 1975, Pasechnik, 1976]. But there was even some minor damage at places the "Tolstyi mys" and the "Troinaya guba" (which are even farther than Irkutsk relative to the epicenter ).

For comparasion in 2014 an earthquake took place about ~250 km to the south of the Tunguska epicenter [Seredkina et al., 2015]. Its moment magnitude was Mw =4.3 (and even larger mb=4.6 ), and a hypocentral depth 6 km. So it was not far in energy from the assigned to the 1908 Tunguska one. It was felt only within ~ 300 km from the epicenter in agreement with the equation.

A proposal that in the place of the earthquake witness's report there were some local geological factors which strongly intensify the ground oscillations helps a little, as the reports of earthquake came from many far-away places. Moreover, if in some place these factors strongly intensify the ground oscillations somehow, then these oscillations would be in the form of very slow swings of the ground. It is because the low-frequency character of the propagating long-distance seismic waves. According to [Ben-Menahem, 1975] the waves at Irkutsk had periods starting from about 8 sec. and towards even lower frequencies. Regarding the relative high-frequency body seismic waves Pasechnik wrote [Pasechnik, 1976] that the waves could be felt by witnesses at the distance from the Tunguska epicenter about several dozens kilometers. So there will be no jerks/pushes and moreover no shivering of the ground at large distances.

Air-waves also can't be responsible for the reported phenomena, as the seismogram at the Magnetic-meteorological observatory in Irkutsk clearly shows that the ground oscillations caused by the airwaves coming at such distances were of low amplitude and of very low frequency [Ben-Menahem, 1975, Pasechnik, 1976]. For example, according to [Ben-Menahem, 1975]: "The periods of the recorded acoustic waves by the Irkutsk seismometer (Fig. lA) vary from 100 to 225 sec.".

Ben-Menahem [Ben-Menahem, 1975], and Pasechnik [Pasechnik, 1976] also stated that durations of the seismic recordings at Irkutsk was up to about 1.5 hours. The long duration of the recording is a puzzle for a long time, as a seismic signal from a spacebody explosive infall must be of relatively short duration. Whipple (who did not see the original recording) wrote in [Whipple, 1930]: "A curious feature of the seismological record is the long duration of disturbance at Irkutsk". He also questioned: could it be a misprint? Later, when it was confirmed, that it was not a misprint, the question was practically ignored (or just assigned to a weak attenuation of seismometers as done in [Pasechnik, 1976]). Indeed Pasechnik [Pasechnik, 1976] wrote (translated, $T_0$ is the period of the free oscillations of the seismograph pendulum):

*"The absence of attenuation in the seismograph explains the significant (1 h 46 min) duration of recording seismic oscillations with a constant period T (20-25 s) equal to $T_0$ at the Irkutsk station (north—south component)."*

Of course, "absence of attenuation" is exaggeration. Ben-Menahem [Ben-Menahem, 1975] considers Q-factor to be about 7-30. This means that with the period of free oscillations 20-30 sec. the amplitude of the "Tunguska earthquake" recordings has to fall off to approximately 4% (i.e. to almost none-detected level on the recordings) for no more than about 15 minutes.

But could the seismograph in the Irkutsk observatory have some unusually low level of damping? The reply is "no". Indeed here is what Voznesenskii wrote [Voznesenskii, 1925] (translated):

*"Since usually the oscillations of the soil fade away quickly during local tremors, then the significant duration of this earthquake is strange, …"*

The author of this paper can confirm this statement. The author of this paper has looked through the Bulletin of the Permanent Central Seismic Commission of Russia for 1908, and other similar documents, and have discovered that durations of recordings by the Irkutsk seismic station of many even much more intensive (in Irkutsk) earthquakes were less than about 20 minutes in that period of time. So the long duration is a puzzle still…

The discrepancy between the testimony of a group of reliable eyewitnesses with the data of records of seismic stations hints at the presence of additional local sources of seismic waves. And there are evidences to this assumption. Here is from [Vasil'ev et al., 1981] (translated):

*"M. Eremenko on July 7, 1908 reported that according to one peasant from the village of Irgey, Nizhneudinsky district. "..During lunch, his hostess staggered with a bowl in her hands to the north. A strong rumble was heard towards the south, like a strong thunderstorm. This phenomenon was repeated twice. Some of the objects hanging on the walls were shaken. It was during lunch, about 12 o'clock in the afternoon." "*

The settlement Irgey is 706 km from the epicenter at azimuth 190° . There is one more similar accoumt from [Vasil'ev et al., 1981]. A paramedic from the village Manzurka ( 858 km from the epicenter at azimuth 161° ) K.S. Sergeev wrote in a letter of July 4, 1908 to Voznesenskii [Vasil'ev et al., 1981] (translated):

*"*
*On Tuesday, June 17 (according to old st. {i.e. the Julian calendar – A.O.}.) About 12 o'clock in the afternoon (pocket unverified watches showed at this moment 11 hours 43 minutes)  during my usual work - reception of outpatients in the Manzurka's district clinic, in which, due to the good sunny weather, most of the windows were opened - a rumble was heard (by me personally) like a shot from a large-caliber cannon, for example, a shot from a siege fortress cannon. This rumble was*

*accompanied by a short insignificant wave-like shaking of the soil, having a direction (as it was possible to determine at that moment) from east to north. In the evening of the same day, the peasant G. Pershin, who came from the field, informed me that around noon (by the sun), while on field work ... about two versts to the north-east from the village Manzurka, during a completely cloudless sky he heard a hum {or a rumble? – A.O.} under the earth with a slight push of the soil.*

    *This message by Pershin, in my opinion is quite plausible and is credible."*

    In addition to his letter K.S. Sergeev also fill the questionnaire B. Here are his replies [Vasil'ev et al., 1981] (translated):

"June 17, 1908 Tuesday.

1. The village of Manzurka, Verkholenskii district, Irkutsk province.
2. Around 12 p.m. on June 17th, 1908.
3. From east to north.
4, 5, 6. An underground rumble was heard with a slight shaking of the soil, which is obtained from a shot from large-caliber artillery cannons.
7. It was not noticed.
8. Sergeev estimated the earthquake with a level of 4 (this figure is underlined)."

    Evidences of a "much later" time (i.e. about 10-12 hours) are also present in the accounts collected in the early 1920s and later. However, due to their lower reliability, they are not represented here.

    But there is also an instrumental  evidence. Pasechnik in [Pasechnik, 1976] published a photocopy of the recording at the Irkutsk seismic station, and here is its scan - see pos. "a" on Fig.4 (unfortunately original printing is of a poor quality).

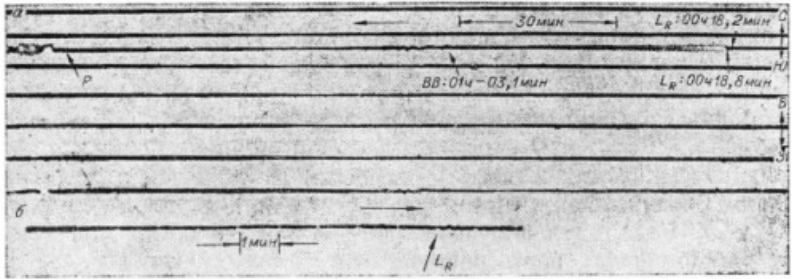
**Fig.4**

On the recording Pasechnik marked pos. "P" as arrival of longitudinal waves of a local earthquake. This took place about 2 hours later after arrival of the "Tunguska earthquake" waves. By the way, please pay attention that duration of the local earthquake recording is much shorter than of the "Tunguska earthquake" – one more proof of the puzzle…

According to Pasechnik [Pasecnhik, 1986] it is not possible to discover seismic waves from other possible (weaker) seismic sources which reached the Irkutsk seismic station 12-33 second later after the main one (Pasechnik researched a possibility of a seismic detection of possible "Tunguska spacebody" fragment's falls).

So evidences point to existence to a swarm of minor local earthquakes in the region of the Tunguska event manifestations. By the way the latter allows to return to the report by Peter Sukhodaeff from Stepanovski Mine about phenomena at June 29, 23 h. 43 m. GMT (see early in the text). The existence of such precursor can't be rule out. Indeed there are other arguments for precursors. One of them was published [Bronshten, 1997] by a prominent Soviet/Russian astronomer Vitaly Bronshten. Despite that the account is from "second-hands", Dr. Vitaly Bronshten was known as a careful fact-checker for his own works, so if he decided to publish some fact then this means that he considers it as reliable. The event took place in the town of Kirensk ( 492 km from the epicenter at azimuth 132° ) on the Lena river. Here is how Bronshten describes the event [Bronshten, 1997] (translated):

"

*Ivan liked to get up early and do jogs in one verst. June 30, 1908 morning was not an exception. This morning was cloudless, the sun brightly shone, no any wind. Suddenly Ivan's attention was drawn by the amplifying noise proceeding as it seemed to him, from southeast side of the sky. Neither from the East, nor from the North, nor from the West nothing similar was felt. The sound came nearer. "All this began. - Ivan Suvorov wrote, - on my watches verified the day before by post-office of Kirensk, at 6 hours 58 minutes local time. Gradually coming source of noise began to be listened from South-South-West side and passed into the West-North-western direction that coincided with the shot-up fiery column up at 7 hours 15 minutes in the morning".*

Ivan Suvorov made this record on fields of the illustrated Bible which used in a family. In 1929 - 1930 when Komsomol members - atheists started walking homes and to withdraw religious literature, Agrippina Vasilyevna herself threw the precious Bible into fire. So Ivan Suvorov's records died.

And nevertheless they were not gone - they remained in memory of his son, Konstantin Suvorov many times reading the story of the father and then restored it.

<...>

What surprises us in these accounts? First of all, time of the beginning of audibility of an abnormal sound - 6 hours 58 minutes while the fiery column shot up, in full consent with other definitions, at 7 hours 15 minutes. The Tunguska bolide could not fly, making a sound, for 17 minutes. During this time at a speed of 30 km/s it would fly by 30000 km, that is at 6 hours 58 minutes it was far outside the atmosphere and could not make any sounds. It means, this moment belongs not to the beginning of emergence of a sound, and to some other event, for example to Ivan's exit from the house.

The correct indication of the moment of explosion forces us to reject also all other possible assumptions: for example, that watches of Ivan lagged behind per day for 17 minutes or that local time of Kirensk strongly differed from local times of other points. Moreover, - in the same Kirensk the director of a meteorological station G. K. Kulesh recorded according to indications of a barograph arrival of an air wave (i.e. the same sounds) after 7 hours.

So inexact Ivan defined and the direction from where the sounds came. The Tunguska bolide flew by, by the most exact definitions, to the North from Kirensk. The closest point of a trajectory was from it to the northeast. Then the bolide moved to the North and, at last, to the northwest.

According to Ye. L. Krinov in his book "Tunguska Meteorite" (Moscow: USSR Academy of Sciences, 1949, p. 54) many eyewitnesses later claimed that they heard the sound before they saw the bolide (which in fact could not be). Apparently, this is some kind of property of inexperienced observers who reported what they saw much later, several years after the event.
"

The Bronsten's attempts to "improve" the account will be not commented, just adding that a barograph in Kirensk recorded arrival of the air-pressure disturbance at 7 h. 48 mi, and the barograph detects air-pressure disturbances of very low frequency.

Krinov indeed noticed existence of these precursors in accounts and wrote [Krinov, 1949] (translated):

"... , attention is also drawn to the indication of the sounds that preceded the appearance of the bolide. It turns out that this strange feature is noted by many witnesses, independently of each other."

Krinov admitted that the precursors which manifested as "blows of considerable force" may be a special type of electrophonic phenomena caused by a large scale of the event, or by eyewitness's confusion [Krinov, 1949]. The first explanation is unlikely as can be seen on example of 2013CME – see (Ol'khovatov, 2021a). The second Krinov's idea is universal explanation which can explain almost everything. However it should be remembered that the idea of Tunguska as a spacebody infall is based on eyewitness's accounts.

By the way Krinov also marked in his book [Krinov,1949] a long duration of the sound phenomena in some places ( up to ~1 hour and even longer). This is many times longer than in 2013CME - see also (Ol'khovatov, 2021b).

In the interpretation of the 1908 Tunguska event as a spaceimpact infall, the source of seismic waves is a relatively short-term and localized in space impulse generated by the action of air shock waves on the earth's surface. As follows from the above, this interpretation is not consistent with the eyewitness's accounts. But the accounts are consistent with the idea that there was a swarm of minor earthquakes in the region at those times, especially as there are signs of activation of tectonic processes around June 30, 1908 in the region [Ol'khovatov, 2003].

It is not clear whether the main event near the epicenter triggered the swarm of weak local earthquakes, or this main event was just a part of the swarm (in the frame of the geophysical interpretation of the Tunguska event). So the problem of these precursors is not closed (as well as the Tunguska event as a whole).

# 5. Conclusion

It can be seen from the written above that while the 2013CME is in a row of typical meteoritic events, but Tunguska strongly deviates from it. Seismic manifestations are very complex, and there are evidences of swarm of earthquakes. In the author's opinion the reported seismic phenomena (taking together with various geophysical peculiarities at that time [Ol'khovatov, 2003] ) are in favor of geophysical interpretation of Tunguska ( more info is on the author's web-page http://tunguska.eu5.org/tunguska.htm ). Anyway there are still many aspects of Tunguska which are needed to be researched.

**ACKNOWLEDGEMENTS**
The author wants to thank the many people who helped him to work on this paper, and special gratitude to his mother - Ol'khovatova Olga Leonidovna (unfortunately she didn't live long enough to see this paper published...), without her moral and other diverse support this paper would hardly have been written.